\documentclass[fleqn,usenatbib]{mnras}

\usepackage{newtxtext,newtxmath}
\usepackage[T1]{fontenc}

\DeclareRobustCommand{\VAN}[3]{#2}
\let\VANthebibliography\thebibliography
\def\thebibliography{\DeclareRobustCommand{\VAN}[3]{##3}\VANthebibliography}

\usepackage{graphicx}	% Including figure files
\usepackage{amsmath}	% Advanced maths commands
\usepackage{amssymb}	% Extra maths symbols
\usepackage{multicol}
\usepackage{ulem}
\usepackage{lipsum}
\usepackage[english]{babel}
\usepackage{color}
\usepackage[dvipsnames]{xcolor}

\def \kms{\ifmmode{~{\rm km\,s}^{-1}}\else{~km~s$^{-1}$}\fi}

\title[The post-CE binary central star of ETHOS~1]{The post-common-envelope binary central star of the planetary nebula ETHOS~1}

\author[J. Munday et al.]{
James Munday,$^{1,2}$\thanks{E-mail: james.munday98@gmail.com}
David Jones,$^{2,3}$\thanks{Email: djones@iac.es}
Jorge Garc\'ia-Rojas,$^{2,3}$
Henri M.J. Boffin,$^{4}$
\newauthor
Brent Miszalski,$^{5}$
Romano L.M. Corradi,$^{6,2}$
Pablo Rodr\'iguez-Gil,$^{2,3}$
\newauthor
Mar\'ia del Mar Rubio-D\'iez,$^{7}$
Miguel Santander-Garc\'ia$^{8}$ and Paulina Sowicka$^{9}$
\\
$^{1}$Astrophysics Research Group, Faculty of Engineering and Physical Sciences, University of Surrey, Guildford, Surrey, GU2 7XH, United Kingdom\\
$^{2}$Instituto de Astrof\'isica de Canarias, E-38205 La Laguna, Spain 
\\
$^{3}$Departamento de Astrof\'isica, Universidad de La Laguna, E-38206 La Laguna, Spain\\
$^{4}$European Southern Observatory, Karl-Schwarzschild-str. 2, D-85748 Garching, Germany\\
$^{5}$Australian Astronomical Optics - Macquarie, Faculty of Science and Engineering, Macquarie University, North Ryde, NSW 2113, Australia\\
$^{6}$GRANTECAN, Cuesta de San Jos\'e s/n, E-38712, Bre\~na Baja, La Palma, Spain\\
$^{7}$Centro de Astrobiolog\'ia (CSIC/INTA), 28850 Torrej\'on de Ardoz, Madrid, Spain\\
$^{8}$Observatorio Astron\'omico Nacional (OAN-IGN), Alfonso XII, 3, 28014, Madrid, Spain\\
$^{9}$Nicolaus Copernicus Astronomical Center, Bartycka 18, PL-00-716 Warsaw, Poland
}

\date{Accepted 2020 September 07. Received 2020 September 07; in original form 2020 August 06}

\pubyear{2020}

\begin{document}
\label{firstpage}
\pagerange{\pageref{firstpage}--\pageref{lastpage}}
\maketitle

\begin{abstract}
We present a detailed study of the binary central star of the planetary nebula ETHOS~1 (PN~G068.1+11.0). Simultaneous modelling of light and radial velocity curves reveals the binary to comprise a hot and massive pre-white-dwarf with an M-type main-sequence companion.  A good fit to the observations was found with a companion that follows expected mass-temperature-radius relationships for low-mass stars, indicating that despite being highly irradiated it is consistent with not being significantly hotter or larger than a typical star of the same mass.  
Previous modelling indicated that ETHOS~1 may comprise the first case where the orbital plane of the central binary does not lie perpendicular to the nebular symmetry axis, at odds with the expectation that the common envelope is ejected in the orbital plane.  We find no evidence for such a discrepancy, deriving a binary inclination in agreement with that of the nebula as determined by spatio-kinematic modelling.  This makes ETHOS~1 the ninth post-common-envelope planetary nebula in which the binary orbital and nebular symmetry axes have been shown to be aligned, with as yet no known counter-examples. The probability of finding such a correlation by chance is now less than 0.00002\%.
\end{abstract}


%%%%%%%%%%%%%%%%%%%%%%%%%%%%%%%%%%%%%%%%%%%%%%%%%%

%%%%%%%%%%%%%%%%% BODY OF PAPER %%%%%%%%%%%%%%%%%%

\section{Introduction}

A significant fraction \citep[perhaps even a majority;][]{demarco15,douchin15} of planetary nebulae (PNe) are the product of interactions in close binary stars \citep{miszalski09}.  However, despite their clear importance in understanding PNe \citep{jones17,boffin19}, the nature of those binary interactions and, in particular, the common envelope (CE) phase is still rather poorly understood.  A key step towards furthering our understanding of these systems is to observationally constrain their parameters such that they can be used to confront models.  In recent years, much progess has been made with a number of post-CE central stars having their masses, radii and temperatures derived via simultaneous modelling of light- and radial velocity curves \cite[see e.g.;][]{jones15,hillwig16,hillwig17,jones19,jones20b}, however for the vast majority of the sixty or so known post-CE central stars only their orbital periods are known.% \citep[and, in some cases, this can be errant by a factor of two!][]{manick15}.

ETHOS~1 (PN~G068.1+11.0, $\alpha$=19$^h$16$^m$31.5$^s$, $\delta$=$-$36\degr{}09\arcmin{}48\arcsec{}) was found by \citet{miszalski11} to host a photometrically-variable central star with a period of roughly 0.535 days.  Spectroscopy of the central star revealed the presence of a complex of emission lines typical of irradiation of a main sequence companion by a hot central star, strongly supporting a binary hypothesis for the photometric variability.  The observed amplitude of this variability ($\sim$0.8 mag in the $i$-band), makes that of the central star of ETHOS~1 one of the most extreme irradiation effects known amongst the central stars of planetary nebulae (CSPNe).  Furthermore, the nebula itself displays a bipolar central structure with extended jets which are found to pre-date the central regions \citep[as has been found in other post-CE PNe;][]{mitchell07,boffin12}.  This combination of extreme irradiation and bipolar jets draws clear parallels with another post-CE CSPN -- that of the Necklace nebula \citep{corradi11}, the companion of which has been shown to be a carbon dwarf having been polluted with carbon-enriched material from the nebular progenitor \citep{miszalski13}.  Previous light- and radial-velocity curve modelling efforts by \citet{mitrofanova16} derived an orbital inclination at odds with the nebular inclination \citep{miszalski11} -- indicating that ETHOS~1 could be the first post-CE PN not to follow the theoretically expected correlation between the two, which is a natural consequence of the CE being ejected preferentially in the orbital plane \citep{hillwig16}.

 In this paper, we present a detailed modelling of the post-CE CSPN of ETHOS~1, based on newly acquired spectroscopy and photometry \citep[combined with the photometry presented in][]{miszalski11}, in order to constrain the binary parameters -- including the orbital inclination and thus probe the possible discrepancy between the binary and nebular inclinations. Section \ref{sec:obs} describes the observations and data reduction, while in Section \ref{sec:model} we present the simultaneous light- and radial velocity curve modelling (and comparison to models published in the literature), before concluding in Section \ref{sec:conc}.

\section{Observations}
\label{sec:obs}
\subsection{Photometry}

Time-series photometry of ETHOS~1 in the $i$-band, acquired using the MEROPE camera \citep{merope} mounted on the 1.2-m Flemish Mercator Telescope between August 24 and September 4 2009, was already presented in \citet{miszalski11}.  The same data is reanalysed here in order to provide a longer time baseline with which to constrain the orbital period as well as model the photometric variability.

Further observations were obtained using the Wide Field Camera (WFC) of the 2.5-m Isaac Newton Telescope (INT)  on the nights of August 21, 22 and 23, as well as November 1 2015, and with the IO:O instrument of the 2-m Liverpool Telescope \citep[LT;][]{LT} on multiple nights between April 29 and July 1 2020. 
With both instruments, exposures were acquired through $B$, $g$, $r$ and $i$ filters with integration times of 120s, 90s, 120s, 90s respectively.  For the exact dates of all exposures, we refer the reader to the online data or the relevant online data archives.

The Mercator and INT observations were debiased and flat-fielded using standard routines of the astropy-affiliated python package {\tt ccdproc} \citep{ccdproc}, while those from the LT were reduced with the IO:O pipeline.  Differential photometry of the central stars was then performed against field stars using the \textsc{sep} implementation of the \textsc{SExtractor} algorithms \citep{bertin96,sep}, before being placed on an apparent magnitude scale using observations of standard stars taken during the course of the observations. The resulting photometric measurements are available in the online data.

\begin{figure*}
\centering
\includegraphics[width=\textwidth]{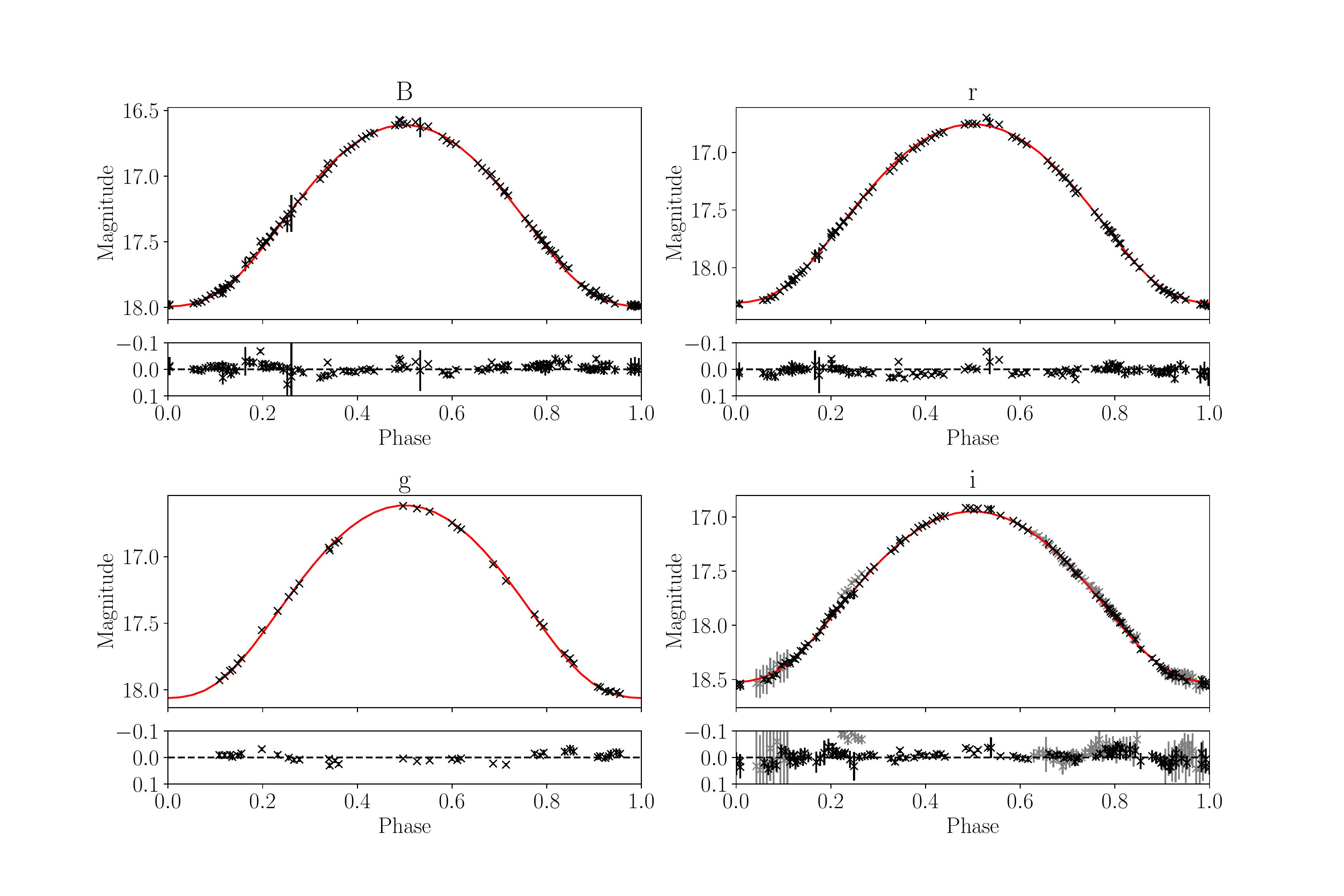}
\caption[]{Phase-folded multi-band photometry of the central star system of ETHOS~1 overlaid on the best-fitting \textsc{phoebe}2 model light curves. The Mercator-Merope $i$-band data, originally presented in \citet{miszalski11}, is shown underlaid in grey beneath the more precise LT-IO:O and INT-WFC data.  Beneath each light curve are the residuals between the best-fitting model and the data.
}
\label{fig:phot}
\end{figure*}

The $i$-band light curve (which offers the longest time span of observations) was searched for periodicities using the \textsc{period} package of the \textsc{starlink} software suite \citep{period,starlink}.  The determined ephemeris is 
\begin{equation}
\mathrm{HJD}_\mathrm{min}=245\,5076.0352(4) +  0.5351263(2) E
\end{equation} 
for the Heliocentric Julian Date of the photometric minimum (HJD$_\mathrm{min}$).  The refined period derived here is entirely compatible with that found by \citet{miszalski11}, however the new, more precise LT and INT data allow us to derive a more accurate timing for the photometric minimum \citep[which is roughly 5$\sigma$ from the value found by][]{miszalski11}.  The phase-folded light curves for each filter are shown in figure \ref{fig:phot}.

The folded light curves all display roughly sinusoidal variability with amplitude increasing with the effective wavelength of the filter \citep[from semi-amplitude $\sim$0.7mag in $B$ to $\sim$0.8mag in $i$, well aligned with the amplitudes found by][]{mitrofanova16}.  As highlighted by \citet{miszalski11}, the large amplitude of variability places the central star of ETHOS~1 among a small group of very extreme irradiation effect binaries \citep{miszalski09,exter03,corradi11}, all of which are found inside PNe. While the amplitude of variability is found to increase with effective wavelength of the filter, the brightness at minimum is found to decrease (reddened $B-r\approx-0.35$ and $r-i\approx0.19$).  This behaviour is generally consistent with the presence of a very hot central star irradiating a main-sequence companion.

\subsection{Spectroscopy}

The central star system of ETHOS~1 was observed using the blue-arm of the Dual Imaging Spectrograph (DIS) mounted on the 3.5m Astrophysical Research Consortium (ARC) telescope at the Apache Peak Observatory on the nights of June 21, 23 and 24 2012 (for exact times, please see table \ref{tab:RVs}).  A 1.5'' wide longslit was employed in conjunction with the B1200 grating to provide a resolution of $\sim$1.8\AA{} between 3900--5050\AA{}.  All exposures were of 1800-s duration and taken at a position angle of 149$^\circ$ (along the jet axis).  The data were debiased, flat-field corrected, optimally extracted and wavelength calibrated using standard \textsc{iraf} routines\footnote{IRAF is distributed by the National Optical Astronomy Observatories, which are operated by the Association of Universities for Research in Astronomy, Inc., under cooperative agreement with the National Science Foundation.}.

The central star system of ETHOS~1 was also observed using the blue-arm of the Intermediate-dispersion Spectrograph and Imaging System (ISIS) mounted on the 4.2-m William Herschel Telescope (WHT) on the nights of August 7 2015, September 2 and 3 2016 (for exact times, please see table \ref{tab:RVs}).  A 1'' wide longslit was used along with the R1200B grating to provide a resolution of $\sim$0.8\AA{} over the range of roughly 4250--5050\AA{}.  All exposures were of 1800-s duration and taken at the parallactic angle.  The data were debiased and flat-fielded using day-time calibrations and wavelength calibration against CuNe+CuAr arc lamp frames (taken either immediately before or after each observation).  The data were then sky-subtracted and optimally extracted \citep[following the algorithm of][]{horne86} to produce 1-D spectra of the central star, all using standard \textsc{starlink} routines \citep{figaro,starlink}. 

\begin{table}
\caption{Heliocentric radial velocity measurements of the irradiated emission line complex in the spectrum of the central star of ETHOS~1.  The entry marked with an asterisk corresponds to the single measurement made by \citet{miszalski11} based on their VLT-FORS2 spectrum.}
\centering
%\begin{center}
\begin{tabular}{ r r r l}
\hline
HJD & Phase & \multicolumn{2}{c}{RV (km~s$^{-1}$)} \\
\hline
$^*$2455026.69807 & 0.80 & $-$122.0 & $\pm$4.6\\
%\hline
2456099.89722 & 0.31 & 91.1 & $\pm$15.0\\
2456099.92182 & 0.36 & 84.9 & $\pm$25.7\\
2456101.90866 & 0.07 & 59.9 & $\pm$33.5\\
2456101.92978 & 0.11 & 61.2 & $\pm$24.4\\
2456103.89864 & 0.79 & $-$104.2 & $\pm$43.5\\
%\hline
2457242.40598 & 0.34 & 111.5 & $\pm$10.3 \\
2457242.42795 & 0.38 & 82.9 & $\pm$13.3 \\
2457242.45300 & 0.42 & 63.8 & $\pm$14.8 \\
2457242.47305 & 0.46 & 43.5 & $\pm$34.0 \\
2457634.39103 & 0.85 & $-$97.7 & $\pm$8.4 \\
2457634.41556 & 0.89 & $-$70.5 & $\pm$23.7 \\
2457634.52229 & 0.09 & 31.9 & $\pm$13.0 \\
2457634.54623 & 0.13 & 69.8 & $\pm$11.7 \\
2457634.56991 & 0.18 & 98.9 & $\pm$29.4 \\
2457634.59331 & 0.22 & 106.7 & $\pm$28.0 \\
2457635.37578 & 0.69 & $-$127.1 & $\pm$5.9 \\
2457635.39946 & 0.73 & $-$131.2 & $\pm$22.0 \\
2457635.42283 & 0.77 & $-$130.1 & $\pm$24.2 \\
2457635.47723 & 0.88 & $-$92.3 & $\pm$6.2 \\
2457635.50076 & 0.92 & $-$64.9 & $\pm$25.6 \\
\hline 
\end{tabular}
\label{tab:RVs}
%\end{center}
\end{table}

\begin{figure}
 \includegraphics[width=\columnwidth]{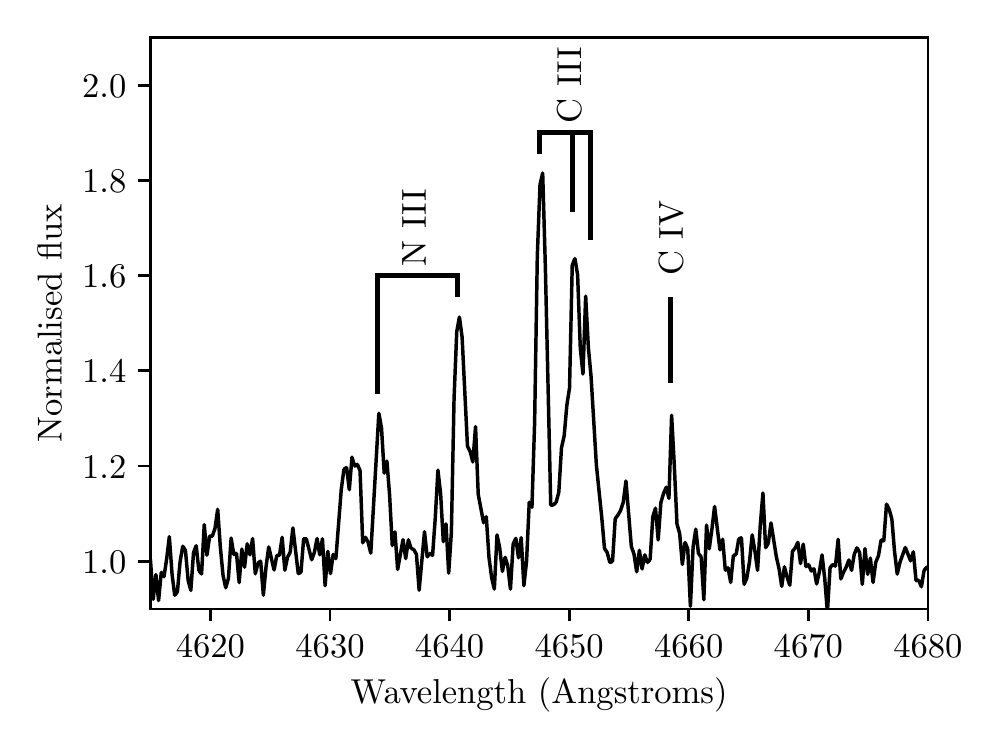}
 \caption{Normalised spectrum of the central star of ETHOS~1 (taken at phase $\phi=0.34$) showing the irradiated emission line complex used to derive the radial velocities of the companion.}
 \label{fig:irrad}
\end{figure}

Both the WHT-ISIS spectra and ARC-DIS spectra were then continuum subtracted before cross-correlation against a template comprising a flat continuum with the complex of irradiated emission lines (N~\textsc{iii} ${\lambda}$4634.14+4640.64\,\AA{}, C~\textsc{iii} ${\lambda}$4647.42+4650.25+4651.47\,\AA{} and C~\textsc{iv} ${\lambda}$4658.30\,\AA{} as identified in figure \ref{fig:irrad}) superimposed \citep[as in][]{jones20b}.  Unfortunately, just as in \citet{miszalski11}, it was not possible to adequately subtract the bright nebular emission in order to derive the radial velocities (RVs) of the hot component of the binary.

\begin{figure}
\centering
\includegraphics[width=\columnwidth]{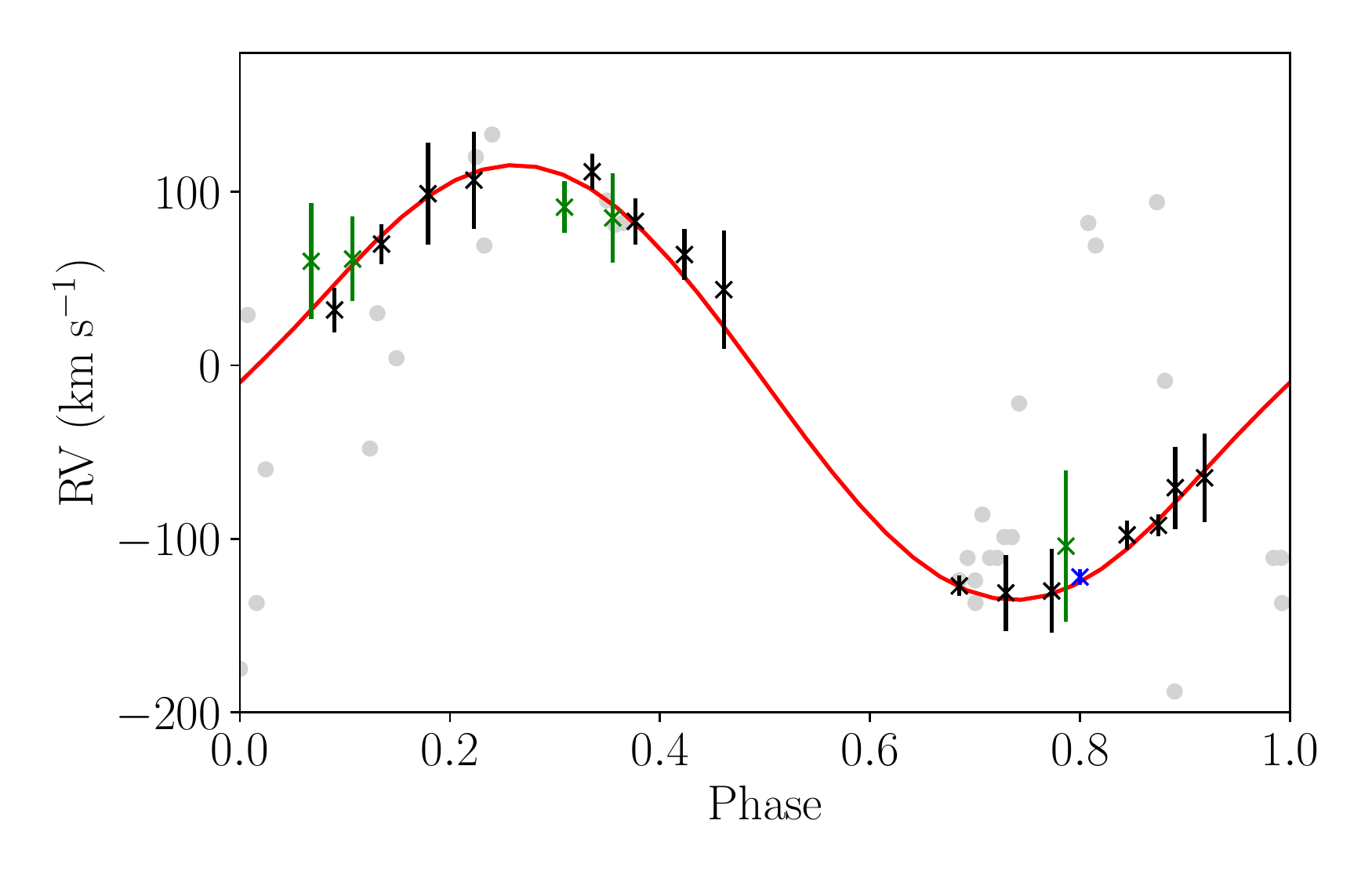}
\caption[]{Phase-folded RVs of the irradiated emission line complex from the central star of ETHOS~1 \citep[WHT-ISIS in black, ARC-DIS in green, and the single VLT-FORS2 point in blue taken from][]{miszalski11}, overlaid on the best-fitting \textsc{phoebe}2 model RV curve.  The radial velocity measurements of \citet{mitrofanova16}, offset by $-105$~km~s$^{-1}$ and folded on the ephemeris determined in Section \ref{sec:obs}, are shown as grey circles (note that they do not quote any uncertainties).  
}
\label{fig:RV}
\end{figure}

The RV measurements of the irradiated emission line complex, following heliocentric correction, are shown in table \ref{tab:RVs} while the data are shown folded on the ephemeris determined from the photometry in Figure \ref{fig:RV}.  Upon inspection, the radial velocities present with a roughly sinusoidal variation with amplitude $\sim$125\kms{} and a systemic velocity $\sim-$10~km~s$^{-1}$.  This amplitude is in good agreement with that determined by \citet{mitrofanova16}\footnote{Note that to derive the RV amplitude, \citet{mitrofanova16} claim to discard all measurements between phases 0.79--1.25 based on the reduced strength of the emission-line complex around photometric minimum. However, it seems clear from their figure 4 c.f. their table 1 and our Figure \ref{fig:RV}, that they also discard at least two data points outside}, however the systemic velocity is rather different.  \citet[][based on their figure 2]{mitrofanova16} find a systemic velocity of approximately 105\kms{}, while the systemic velocity found here is much closer to that derived for the nebula, $v_\mathrm{sys}\sim-$20\kms{}, by \citet{miszalski11}.  It is also interesting to note that the systemic velocity implied by our measurements and those of the nebula are roughly consistent with Galactic rotation (accounting for a correction to the Local Standard of Rest), while the velocity of \citet{mitrofanova16} would likely indicate some form of runaway star.

\section{PHOEBE modelling}
\label{sec:model}

The extracted light and radial velocity curves were modelled using the version 2.2 release of the \textsc{phoebe}2 code \citep{prsa16,horvat18,jones20}. Fitting was performed via a Markov Chain Monte Carlo (MCMC) method in python, using the \textsc{emcee} package, and executed on the LaPalma3 supercomputer \citep[using the methodology presented in][]{jones19}.

The primary's temperature, mass and radius were allowed to vary freely, with limb-darkening coefficients extrapolated from the tables of \citet{claret20}. \textsc{phoebe2}'s interpolated limb-darkening, based on \citet{ck2004} atmospheres, were used for the secondary, while its albedo was allowed to vary freely in the physically-acceptable range 0.6--1.0 \citep{claret01}.  As there are minimal observational constraints for the secondary's mass (i.e.\ no radial velocities from the primary component) and radius (the system is not eclipsing), the mass, temperature and radius of the secondary were reduced to a single free parameter (the mass) with the others being set using the mass-temperature-radius relationship (MTRR) found for main-sequence stars by \citet[][note that there is a typographic error in their mass-radius relationship for the low mass range\footnote{The correct mass-radius relationship for the range 0.179$\leq$M/M$_\odot$$\leq$1.5, as kindly provided by Prof.~Z.~Eker, should be: \hbox{R= 0.438(098) $\times$ M$^2$ + 0.479(180) $\times$ M + 0.137(075)}.} which was discovered by our referee, Prof.~Todd Hillwig]{eker18}.  The final free parameters in the model are the binary inclination, allowed to vary freely over a range ($35^\circ<i<70^\circ$) which covers both the solutions of \citet[][$37^\circ\pm1^\circ$ and $39^\circ\pm0.2^\circ$]{mitrofanova16} and the inclination expected from spatio-kinematical modelling of the nebula \citep[][$60^\circ\pm5^\circ$]{miszalski11}, and the systemic velocity (allowed to vary over the range, $\gamma=[-$30, $+$10]~km~s$^{-1}$).

\begin{figure*}
\centering
\includegraphics[width=0.98\textwidth]{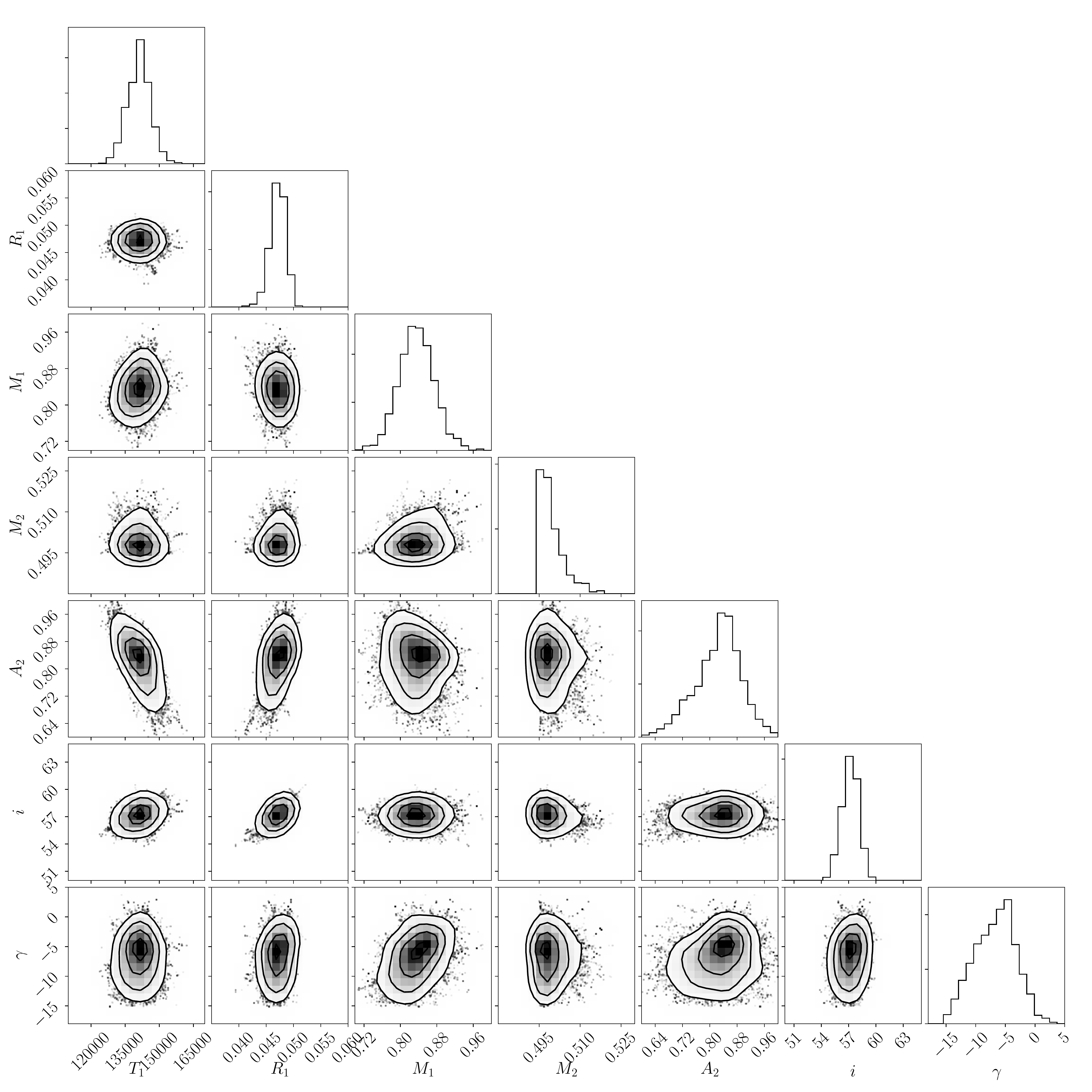}
\caption[]{A corner plot \citep[made using \textsc{corner};][]{corner} of the \textsc{phoebe}2 MCMC posteriors for the primary temperature ($T_1$), radius ($R_1$) and mass ($M_1$), and secondary mass ($M_2$) and albedo ($A_2$), as well as the binary orbital inclination ($i$) and systemic velocity ($\gamma$).}
\label{fig:ETHOS1mcmc}
\end{figure*}

The best-fitting model parameters (as derived from sampling the MCMC posteriors as shown in Figure \ref{fig:ETHOS1mcmc}) are listed in Table \ref{tab:params}, while the corresponding model light- and RV-curves are shown underlaid on the data in Figures \ref{fig:phot} and \ref{fig:RV}, respectively.  

The model fits the observed spectroscopy and photometry well, with residuals generally being of the order of one uncertainty.  The majority of photometric points where this is not the case, are likely a consequence of underestimated uncertainties as, while every attempt has been made to remove the nebular contamination, this is not reflected in the purely statistical uncertainties. The model residuals are, however, systematically larger than the uncertainties around the $i$-band minimum, perhaps indicative that the temperature of the secondary is over-estimated by the model (this is the only band where the companion will contribute significantly around photometric minimum due to the high temperature and luminosity of the hot component).  Relaxing the MTRR could potentially resolve this discrepancy, but given the lack of additional constraints and the intrinsic problems involved in modelling irradiated binaries \citep[see section 7 of][for a more detailed discussion]{horvat19}, we feel that the forced-MTRR fit is reasonable.  Similarly, the choice of model atmosphere for the secondary \citep[][as is standard in \textsc{phoebe}2]{ck2004} may also be important with the derived secondary temperature lying so close to the lower boundary of the range covered by those atmospheres (the shape of the MCMC posteriors for $M_2$ in Figure~\ref{fig:ETHOS1mcmc} give some indication that this lower limit is hit by a number of chains).  In spite of this, we still favour these atmospheres over the other options: \textsc{phoenix} \citep[the range of which does not extend to high enough temperatures to model such extreme irradiation][]{husser13} or black body atmospheres (which are a rather poor representation of the observed spectral energy distribution of low-mass dwarfs). It is important to note that forcing a lower inclination, similar to that found by \citet{mitrofanova16}, does not improve the fit, instead leading to a larger discrepancy around the $i$-band minimum.

\begin{table}
\caption{Best-fitting model parameters for the central star.  Parameters marked with an asterisk were derived using the MTRR of \citet{eker18}.}
\centering
%\begin{center}
\begin{tabular}{ r r l}
\hline
Parameter & \multicolumn{2}{c}{Best fitting value} \\
\hline
Primary mass (M$_\odot$) & 0.84 & $\pm$0.04 \\
Primary temperature (kK) & 141 & $\pm$ 6\\
Primary radius (R$_\odot$) & 0.047 & $\pm$0.002 \\
Secondary mass (M$_\odot$) & 0.50 & $\pm$ 0.01\\
Secondary temperature$^\star$ (kK) & 3.5 & $\pm$ 0.1\\
Secondary radius$^\star$ (R$_\odot$) & 0.48 & $\pm$ 0.01\\
Secondary albedo & 0.83 & $^{+0.06}_{-0.08}$\\
Binary inclination ($^\circ$) & 57.3 & $^{+0.8}_{-1.0}$\\
Systemic velocity, $\gamma$ (km~s$^{-1}$) & $-$6.5 & $^{+3.2}_{-4.0}$\\
\hline 
\end{tabular}
\label{tab:params}
%\end{center}
\end{table}

\subsection{Discrepancies with the modelling of \citet{mitrofanova16}}
\label{sec:mitro}

\citet{mitrofanova16} presented a model of the CSPN of ETHOS~1 based on three-band photometry and radial velocity measurements of the irradiated emission line complex originating from the secondary.  Those authors find fits consistent with the central star being an extremely hot white dwarf with a mass of either 0.7 or 0.78~M$_\odot$.  
However, perhaps due to the lack of observations between phases 0.4--0.7 where the emission lines are strongest, less than half of their spectra could be used to derive the radial velocity curve (see Figure \ref{fig:RV} c.f.\ their figure 4).  In deriving the mass function of the binary, they include a K-correction in order to account for the fact that the emission line complex is not reflective of the dynamical (centre-of-mass) velocity of the companion \citep[but rather the centre of light velocity which is skewed towards the first Lagrange point due to the irradiation from the hot primary;][]{exter03,miszalski11}.  However, for such a large irradiation effect, their correction of only 7--9\kms{} seems rather small -- for example, applying the equation of \citet{wood95}, which derives the K-correction assuming that all parts of the irradiated hemisphere contribute equally to the emission while the outward facing hemisphere does not contribute at all, would imply a K-correction of $\sim$15\kms{} for their model parameters.  Note that this is likely the minimum correction as not all parts of the irradiated hemisphere will contribute equally, with the most irradiated regions (i.e. those closest to the primary) contributing more.  Thus, the potentially underestimated K-correction could, if the orbital inclination of the model holds (which our modelling indicates is likely not the case), lead to a significantly underestimated mass for the primary star \citep{exter03}.

It is unclear how \citet{mitrofanova16} derive the K-correction, other than that it is apparently based on synthetic spectra. However, it clearly does not correspond to the centre-of-light of the irradiated hemisphere (which is the approach taken in \textsc{phoebe2}, roughly equivalent to a K-correction of $\sim$25\kms{} for their model parameters).  For the K-correction of \citet{mitrofanova16} to be valid, the irradiated emission lines would have to originate from the entire surface of the companion (even the non-irradiated hemisphere) or not be produced in the most irradiated regions of the irradiated hemisphere (i.e. restricted to a ring/band perpendicular to the orbital plane and displaced slightly towards the first Lagrange point).  Importantly, applying a larger K-correction to the \citet{mitrofanova16} models would imply a primary mass approaching the Chandrasekhar limit and thus an exceptionally massive progenitor (our model has a greater inclination and thus avoids this issue).

\citet{mitrofanova16} did not consider the expected link between nebular inclination and binary orbital plane \citep{hillwig16} and choose to restrict the inclinations of their models to inclinations between 35--45$^\circ$, based on the light curve morphology.  We do allow our model fits to explore inclinations as low as those found by \citet{mitrofanova16}, but also extend the range upwards to encompass inclinations compatible with that of the nebula (the final range explored being $35^\circ < i < 70^\circ$).  Ultimately, we find that our best-fitting model presents with an inclination in good agreement with that of the nebula (57.3$^\circ$$\pm$1.0$^\circ$ c.f.\ 60$^\circ$$\pm$5$^\circ$), strongly indicating that ETHOS~1 does actually follow the expected correlation between binary plane and nebular symmetry axis just as for every other case where this has been tested \citep{hillwig16}.  If we force our models to the inclination found by \citet{mitrofanova16}, the fit becomes appreciably worse (particularly in the $i$-band, in which they did not have data) and the primary mass tends towards the Chandrasekhar limit.

We find that we are able to fit the observed light and radial velocity curves using a model secondary which follows a standard MTRR for main-sequence stars.  The model secondaries of \citet{mitrofanova16}, on the other hand, have much larger radii than predicted using the same MTRR.
Although there is a precedence for secondaries with inflated radii in post-CE binary central stars \citep[e.g.;][]{jones15}, the radii of the \citet{mitrofanova16} model secondaries are very large for their model masses (being more than a factor of 3 and 2 larger than typical for their two model solutions, respectively).  However, it is unclear how their secondary masses are constrained in the modelling (the radial velocities of the primary star are not measured).  Thus, given that we can model the system with a companion which follows a standard MTRR for main-sequence stars, we conclude that there is no evidence of inflation in the secondary of ETHOS~1.  This is a particularly interesting conclusion given that most post-CE central stars are found to present at least some level of inflation \citep{jones15,jones20b}.  However, ETHOS~1 would not be the first such case \citep[e.g.; the central star of M~3-1,][although this may be a special case as the secondary is very close to Roche-lobe filling]{jones19}.  Moreover, studies of more evolved post-CE white-dwarf-main-sequence binaries find that the M-dwarf companions in those systems are indistinguishable from field systems \citep{parsons18}, indicating that such inflation may not be a universal characteristic of the CE or, at the very least, not so long-lived.

Another likely difference in our modelling approach is in the choice of secondary albedo, which was a free parameter in our models (encompassing the physically accepted range of 0.6--1.0).  Unfortunately, \citet{mitrofanova16} do not quote the secondary albedo used in their modelling, however if it were fixed to the theoretical value for fully-convective stars \citep[$\sim$0.6;][]{rucinski69}, then this could mean that a larger primary temperature (as in both their models) or a physically larger companion would be required to replicate the observed photometric amplitudes if the true albedo were higher -- just as we find.  Similar studies of highly irradiated binaries do tend to empirically derive larger albedos \citep{rafert80}, while more-detailed theoretical studies similarly predict larger albedos ($\sim$0.75--0.8, in line with that derived here) at the effective temperature of their model companion \citep{claret01}.

The models of \citet{mitrofanova16} imply a distance \citep[in order to match the observed brightness, assuming the extinction determined by][from the nebular spectrum]{miszalski11} of between 4--6 kpc, while our model would indicate a distance of $\sim$4.2 kpc.  No distance within this range can be discarded based on the observed kinematical ages of both the central regions and jets, which would be 4--10 kyr depending on the model.  The 0.78~M$_\odot$ model of \citet{mitrofanova16} provides the best match to the distance derived using the H$\alpha$-surface-brightness-radius relation of \citet{frew16} at 6.92$\pm$1.26 kpc. However, there is some indication that post-CE PNe are less massive and less luminous than their single-star counterparts \citep{frew08,corradi15}, which would lead to an overestimated distance.  As such, the implied distances do not rule out any particular model \citep[our own or either model of][]{mitrofanova16}.

\citet{mitrofanova16} also chose to make their model primaries lie on the evolutionary tracks of \citet{bloecker95}, which do not take into account the significant advancements in our understanding of AGB star evolution from the last decades \citep{mmmb}.  In any case, these models are for single stars and may not necessarily be entirely representative of post-CE systems \citep{mmmb17,jones19,jones20b}.  
Putting these considerations aside, the best fitting 0.7~M$_\odot$ model of \citet{mitrofanova16} lies on the evolutionary track of \citet{bloecker95} for an age of $\sim$700yr\footnote{No \citet{bloecker95} tracks are available for a remnant mass of 0.78~M$_\odot$, so it is not possible to test the age of the solution of \citet{mitrofanova16} with this mass.  However, interpolation of the higher mass tracks at 0.84~M$_\odot$ and lower mass at 0.7~M$_\odot$ indicate that this solution would have a similarly young post-AGB age.}, inconsistent with the larger kinematical ages required for their models to reproduce the observed angular dimensions of ETHOS~1.

Based on the initial-to-final-mass relation of \citet{cummings18}, the initial mass of the CSPN would have been 2.7--3.3~M$_\odot$ for the remnant masses derived by \citet{mitrofanova16}, and $\sim$3.5~M$_\odot$ for the remnant in our model. All of which are rather massive for the Galactic latitude and estimated distance of ETHOS~1. Single-star nucleosynthesis models \citep[e.g.;][]{ventura13,karakas16} predict that third dredge-up should be activated in stars with initial masses in these ranges, ultimately leading to elevated C and s-process element abundances in the surrounding PN. One may thus look for indicators of the initial mass in the nebular abundances of ETHOS~1.  Unfortunately, the intrinsic faintness and density-bounded nature of the nebula limits the drawing of strong conclusions, with no s-process lines detected and the C/O ratio typically measured using faint lines in the UV (outside the range of the available observations).

As highlighted in \citet{garcia-rojas18}, the He/H ratio can also be a probe of the initial mass -- particularly for more massive progenitors when combined with the N/O ratio.  In the central nebula of ETHOS~1, the He~{\sc ii} $\lambda$4686 line is brighter than H$\beta$, however no He~{\sc i} emission was detected.  Note that He~{\sc i} $\lambda$4471 was identified in the SE jet (although only weakly), but the lines fluxes of the jets are difficult to interpret in terms of chemical abundances due to their shock-excited nature, thus we do not consider them here. Assuming that all He in the central nebula is in the form of He$^{++}$, the obtained abundance is 12+log(He/H)=11.01$\pm$0.04 -- relatively low and perhaps indicative of a low initial mass.  However, there are significant degeneracies between models, depending on the physics assumed and the metallicity \citep{ventura13,karakas16}, meaning that initial masses up to $\sim$4.5~M$_\odot$ could fit with such an abundance.  As such, the nebular abundances \citep[based on the few nebular lines observed by][]{miszalski11}, cannot be used to distinguish effectively between our model and those of \citet{mitrofanova16}.

As already mentioned, the He~{\sc ii} $\lambda$4686 line is brighter than H$\beta$ in ETHOS~1 -- a clear sign that the nebula is density-bounded (i.e. all the nebular material has been ionised).  However, the ionised mass of the nebula is found to be strikingly low at roughly \mbox{$0.002\times D^2$~M$_\odot$}, where $D$ is the distance to ETHOS~1 in kpc \citep[][Santander-Garc\'ia et al., in prep]{santander-garcia19}.  Similarly minuscule masses have been noted for several other post-CE PNe \citep{corradi15}, seemingly at odds with the idea that the nebula represents the majority of the progenitor's envelope (which would be at least an order of magnitude more massive) rapidly ejected during the CE event \citep{corradi14,corradi15}.

\section{Conclusions}
\label{sec:conc}

We have presented a detailed study of the binary nucleus of ETHOS~1, indicating that the central star is a $\sim$140~kK pre-white-dwarf of mass $\sim$0.84~M$_\odot$ -- making it one of the hottest and most-massive post-CE central stars known \citep[see e.g.; the central star of NGC~2392;][]{miszalski19}.    A similar study by \citet{mitrofanova16} found model fits with central star masses of 0.7 and 0.78~M$_\odot$, in relatively good agreement with the mass derived here in spite of finding a significantly lower orbital inclination (the differences between their analyses and the one presented here are discussed at length in Section \ref{sec:mitro}).

The orbital inclination of our model is found to be in good agreement with the nebular inclination as derived via morpho-kinematical modelling \citet{miszalski11}.  This is at odds with the findings of \citet{mitrofanova16}, who found a binary inclination discrepant by roughly 20$^\circ$ -- which would make ETHOS~1 the first post-CE PN not to follow the expected correlation between binary plane and nebular symmetry axis \citep{hillwig16}.  It should also be noted that, assuming the measured radial velocities are representative of the companion's centre of light,  the mass of the primary would be much larger (approaching the Chandrasekhar limit) for the orbital inclination found by \citet{mitrofanova16}.

Although the radius and mass of the companion are relatively poorly constrained in the \textsc{phoebe}2 modelling (due to the binary's non-eclipsing nature and the lack of RV measurements from the primary), a good fit was found using parameters compatible with being a normal main-sequence star.  This could be an indication that, in the case of ETHOS~1, the mass transfer which led to the formation of the jets did not significantly impact upon the physical parameters of the companion \citep[unlike in the majority of well-studied, main-sequence companions to post-CE PN central stars;][]{jones15}.

To conclude, our modelling indicates that the central star of ETHOS~1 is one of the hottest (and most massive) known inside a post-CE PN. As the system is not eclipsing, nor double-lined, the parameters of the secondary are difficult to constrain independently.  However, models using a MTRR find satisfactory solutions indicating that its properties are comparable to typical early M-type stars.  The binary inclination is consistent with lying perpendicular to the nebular symmetry axis, as determined by spatio-kinematical modelling, adding further weight to the empirical correlation found by \citet{hillwig16}.  Indeed, adding ETHOS~1 to the eight systems in \citet{hillwig16}, the likelihood of chance alignment drops to less than 0.00002\%.

\section*{Acknowledgements}

The authors would like to thank the referee, Prof.~Todd Hillwig, for his extremely insightful report which greatly improved the paper. The authors are also grateful to Prof.~Bruce Balick for helping to obtain the APO data, and to Prof.~Z.~Eker for providing the corrected form of the mass-radius relationship.

JM acknowledges the support of the ERASMUS+ programme in the form of a traineeship grant. DJ, JGR, PRG and RLMC acknowledge support from the State Research Agency (AEI) of the Spanish Ministry of Science, Innovation and Universities (MCIU) and the European Regional Development Fund (FEDER) under grant AYA2017-83383-P.  DJ and JGR also acknowledge support under grant P/308614 financed by funds transferred from the Spanish Ministry of Science, Innovation and Universities, charged to the General State Budgets and with funds transferred from the General Budgets of the Autonomous Community of the Canary Islands by the Ministry of Economy, Industry, Trade and Knowledge.  MSG acknowledges support from the Spanish Ministry of Science, Innovation, and Universities under grant AYA2016-78994-P.  PS thanks the Polish National Center for Science (NCN) for support through grant 2015/18/A/ST9/00578.

The authors thankfully acknowledge the technical expertise and assistance provided by the Spanish Supercomputing Network (Red Espa\~nola de Supercomputaci\'on), as well as the computer resources used: the LaPalma Supercomputer, located at the Instituto de Astrof\'isica de Canarias.

Based on observations made with the Isaac Newton Telescope and the William Herschel telescope both operated by the Isaac Newton Group of Telescopes, with the Mercator Telescope operated by the Flemish community, and with the Liverpool Telescope operated by by Liverpool John Moores University with financial support from the UK Science and Technology Facilities Council, all of which reside on the island of La Palma at the Spanish Observatorio del Roque de los Muchachos of the Instituto de Astrof\'isica de Canarias. This research made use of Astropy,\footnote{http://www.astropy.org} a community-developed core Python package for Astronomy \citep{astropy:2013, astropy:2018}.

\section*{Data availability}
All raw LT-IO:O, INT-WFC and WHT-ISIS data are available from the respective online archives, while the Mercator-Merope and ARC-DIS data are available upon reasonable request from the authors.  All extracted photometry and radial velocities are available in the article or from VizieR at the CDS.

%%%%%%%%%%%%%%%%%%%%%%%%%%%%%%%%%%%%%%%%%%%%%%%%%%



\bibliographystyle{mnras}
\bibliography{ETHOS1} % if your bibtex file is called example.bib
\bsp	% typesetting comment

% Don't change these lines

\label{lastpage}
\end{document}